\documentclass[%
reprint,
superscriptaddress,
amsmath,
amssymb,
aps,
prl
]{revtex4-1}

\usepackage{graphicx}% Include figure files

\begin{document}

\title{An Optical Tweezer Array of Ultracold Molecules}

\author{Lo\"ic Anderegg}
\email{anderegg@g.harvard.edu} 
\author{Lawrence W. Cheuk}
\author{Yicheng Bao}
\author{Sean Burchesky}
\affiliation{Department of Physics, Harvard University, Cambridge, MA 02138, USA}
\affiliation{Harvard-MIT Center for Ultracold Atoms, Cambridge, MA 02138, USA}

\author{Wolfgang Ketterle}
\affiliation{Harvard-MIT Center for Ultracold Atoms, Cambridge, MA 02138, USA}
\affiliation{Department of Physics, Massachusetts Institute of Technology, Cambridge, MA 02139, USA }

\author{Kang-Kuen Ni} 
\affiliation{Department of Physics, Harvard University, Cambridge, MA 02138, USA}
\affiliation{Harvard-MIT Center for Ultracold Atoms, Cambridge, MA 02138, USA}
\affiliation{Department of Chemistry and Chemical Biology, Harvard University, Cambridge, MA 02138, USA}

\author{John M. Doyle} 
\affiliation{Department of Physics, Harvard University, Cambridge, MA 02138, USA}
\affiliation{Harvard-MIT Center for Ultracold Atoms, Cambridge, MA 02138, USA}

\date{\today}
\begin{abstract}
Arrays of single ultracold molecules promise to be a powerful platform for many applications ranging from quantum simulation to precision measurement. Here we report on the creation of an optical tweezer array of single ultracold CaF molecules. By utilizing light-induced collisions during the laser cooling process, we trap single molecules. The high densities attained inside the tweezer traps have also enabled us to observe in the absence of light molecule-molecule collisions of laser cooled molecules for the first time.
\end{abstract}

\maketitle

In recent years, various quantum systems ranging from trapped ions and superconducting circuits to ultracold atoms have been brought under exquisite control, opening up many new applications in quantum science. In addition to these well-explored systems, ultracold molecules promise to be a powerful quantum resource~\cite{carr09}. Compared to atoms, they have a much richer internal structure, which gives rise to desirable features such as long-lived states with tunable long-range interactions. These features could be harnessed for creating molecular qubits~\cite{demille02qi,yelin06,karra16,ni18,blackmore18}, quantum simulation of spin lattice Hamiltonians~\cite{zoller06,buechler07,pupillo08}, studies of ultracold quantum chemistry~\cite{krems08,Bala16} and precision measurements that probe physics beyond the Standard Model~\cite{Ye06}.

These promising applications have led to intense efforts to control ultracold molecules, which is challenging due to their complex internal structure. Nevertheless, there has been rapid progress in cooling and trapping of molecules. Starting with the first molecular magneto-optical traps (MOTs)~\cite{barry14,norrgard16RF,truppe17,anderegg17,collopy18}, laser cooled molecules have recently been conservatively trapped~\cite{williams18,McCarron18,anderegg18,cheuk18}, and using atom association the first quantum degenerate molecular gases have been created~\cite{Marco18}. In addition to cooling and trapping, another frontier in controlling atomic and molecular systems is gaining control of  individual particles~\cite{schlosser01,Saffman06,Kaufman12,Kaufman18,cooper18,Saskin18}. Recently, rearrangeable optical tweezer arrays of single atoms have emerged as a powerful bottom-up approach to building quantum systems particle-by-particle~\cite{barredo16array,lukin16array}, including assembly of a molecule from two atoms~\cite{Liu18}. Similar arrays of single molecules, which offer deterministic state preparation and high-fidelity detection, could open up new avenues in quantum simulation and computing.

\begin{figure}[b]
\centering
\includegraphics[scale=.22]{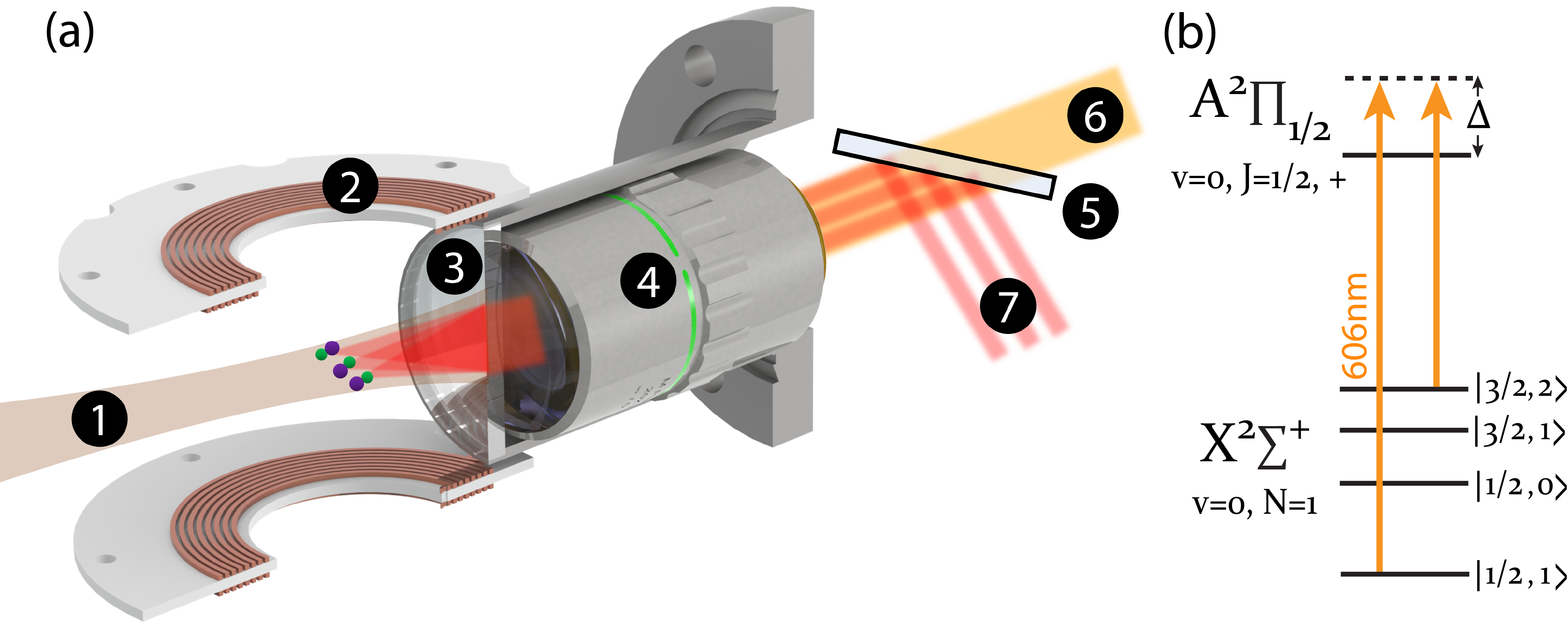}
\label{setup} 
\caption{ \textbf{Molecular Energy Diagram and Experimental Setup} (a) An optical dipole trap formed by a focused beam of 1064~nm light (1) intersects the MOT, and is reflected off the re-entrant window (3) at an angle to prevent the formation of a lattice. A microscope objective (4) is placed inside a re-entrant housing between the MOT coils (2). Fluorescence from the molecules (6) is collected through the objective and imaged onto a camera. The optical tweezer traps are generated using an acousto-optic deflector (AOD) (7) and are combined into the imaging path using a dichoric mirror (5). (b) CaF level structure of relevant states used in the $\Lambda$-cooling process. The cooling is operated at a detuning $\Delta=2\pi\times25$\,MHz. }
\end{figure}

\begin{figure}[b]
\centering
\includegraphics[scale=.42]{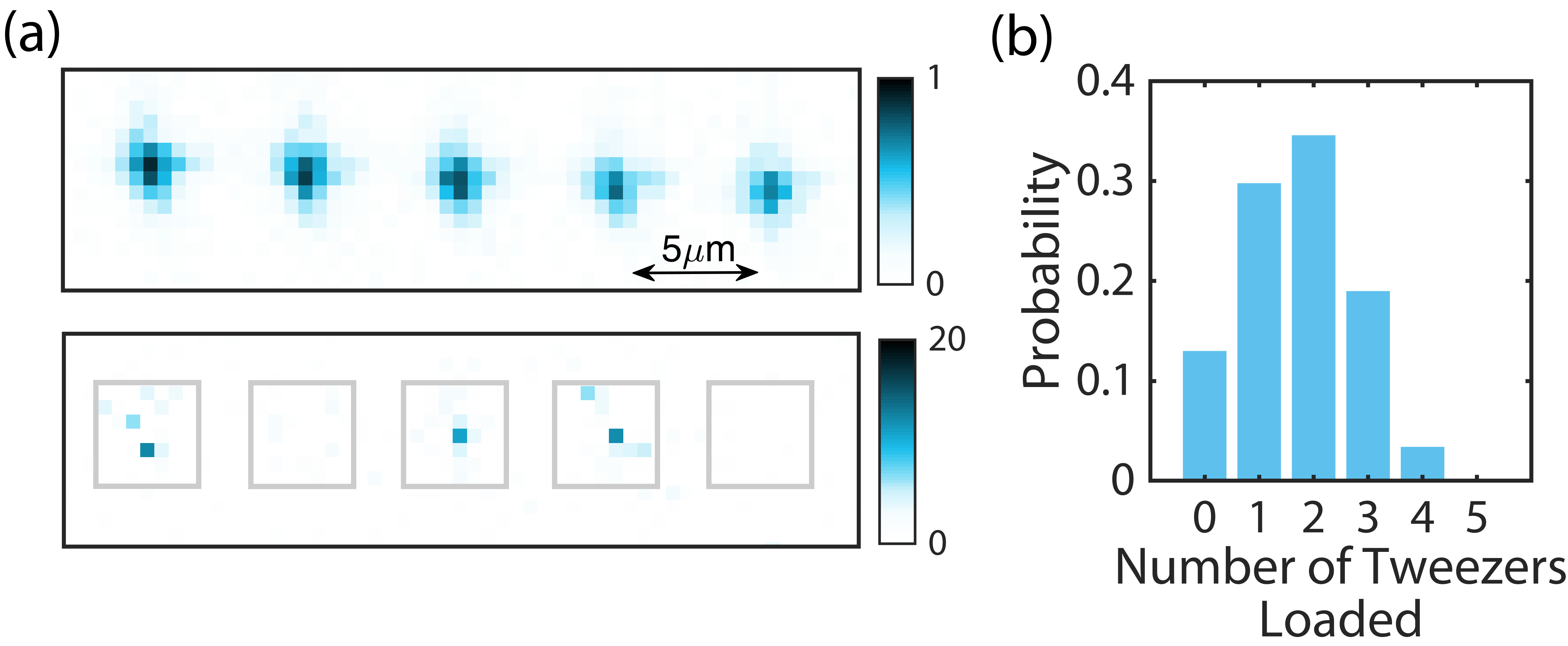}
\label{array} 
\caption{ \textbf{Molecule Tweezer Array.} (a) Top: Image of optical tweezer array of single molecules, averaged over 500 shots. Bottom: Single shot image showing three occupied tweezer traps. The grey boxes represent the regions over which photon counts are summed. (b) Probability versus number of tweezer traps loaded. The average loading probability per trap is 34\%. }
\end{figure}

In this letter, we report successful creation and detection of an array of ultracold calcium monofluoride (CaF) molecules trapped in optical tweezers. CaF molecules are laser cooled into individual optical tweezer traps, during which we directly observe light-assisted collisions, which give rise to collisional blockade~\cite{schlosser01,schlosser02} ensuring single molecule trapping. The measured light-assisted collision rate sets a fundamental density limit for laser cooling molecules in this type of system. In addition, we have achieved molecular densities over an order of magnitude greater than previous experiments, allowing us to observe ground electronic, excited rotational state collisions of laser cooled molecules.

The starting point of our experiment is a MOT of $10^4$ molecules in the ground electronic, vibrational and first rotational (X, $v=0$, $N=1$) manifold~\cite{anderegg17}. The MOT density of $10^5$~cm$^{-3}$ is too low for direct capture into $\mu$m-sized optical tweezers. To reach loading probabilities of order unity, one would require densities of $\sim 10^{11}\,\text{cm}^{-3}$, more than four orders of magnitude higher than the highest achieved MOT density for laser cooled molecules~\cite{anderegg17}. To overcome the low starting densities, we load the optical tweezers using a two-step approach. First, molecules are transferred from the MOT into an optical dipole trap (ODT) formed by a focused 1064\,\text{nm} laser beam (Gaussian beam waist $45\,\mu$m) in the presence of $\Lambda$-enhanced gray molasses cooling ($\Lambda$-cooling) light on the $X\rightarrow A$ transition~\cite{cheuk18}, Fig.~1b. Since the laser cooling continues to work inside the trap, large density enhancements can be obtained~\cite{anderegg18}. The molecules trapped inside the 1064\,nm ODT are subsequently transferred into the smaller $\mu$m-sized optical tweezers (780\,nm) also with the aid of $\Lambda$-cooling.

In detail, after the molecules are loaded into the MOT, the MOT beams and magnetic gradients are switched off, and 100\,ms of $\Lambda$-cooling is applied to load molecules into a $200\,\mu$K deep ODT. This produces trapped samples with densities as high as $3\times10^7$~cm$^{-3}$ at temperatures of $20\,\mu\text{K}$. The trapped molecules are then transferred into the optical tweezer traps, which are formed by tightly-focused 780\,nm laser beams. The optical tweezers are projected through a high-resolution imaging path, created by incorporating a microscope objective into the experiment~(Fig.~1a). Multiple optical tweezers are created using an acousto-optical deflector (AOD). The positions and the depths of the tweezers are controlled by the RF frequencies and powers driving the AOD. Typically, a tweezer trap depth of $330\,\mu$K is used. To transfer molecules from the ODT to the optical tweezer, the $\Lambda$-cooling light is left on while the ODT power is ramped down over a few ms. The ODT and the cooling light is then turned off in order for the remaining molecules to fall away.

The molecules in the optical tweezers are detected via $\Lambda$-imaging~\cite{cheuk18}, where $\Lambda$-cooling is applied and the molecular fluorescence at 606\,nm is collected through the microscope objective and detected on an electron multiplying camera (EMCCD). We find that $\Lambda$-imaging remains effective for the tightly-focused tweezer traps. Using an imaging duration of 30\,ms, $2000$ photons are scattered, of which 30 are detected. Fig.~2a shows an average image of molecules trapped in an array of 5 optical tweezers, as well a typical single shot image with three occupied tweezers. We estimate a loading probability per trap of ~34\% (Fig.~2b) due to the stochastic nature of the loading process and limited initial densities.

To characterize the tweezer traps and the molecular samples, we measure the radial trapping frequency $\omega_r$ via parametric heating, and the molecular temperature via a release and recapture measurement (see Methods). Using the trap frequency and the calculated AC polarizbility of CaF at $780\,$nm, we determine that the tweezer traps have Gaussian beam waists of 2.3\,$\mu$m, in agreement with the input beam size into the objective and measured optical aberrations arising from the re-entrant window. The release and recapture measurements give a temperature of 80(20)\,$\mu$K, 7 times lower than the tweezer trap depth and well-below the Doppler limit ($200\,\mu\text{K}$), verifying that $\Lambda$-cooling remains effective inside the tweezer traps. For two molecules, the peak trapped density is $3\times10^{10}$\,cm$^{-3}$.

\begin{figure}[t]
\centering
\includegraphics[scale=.52]{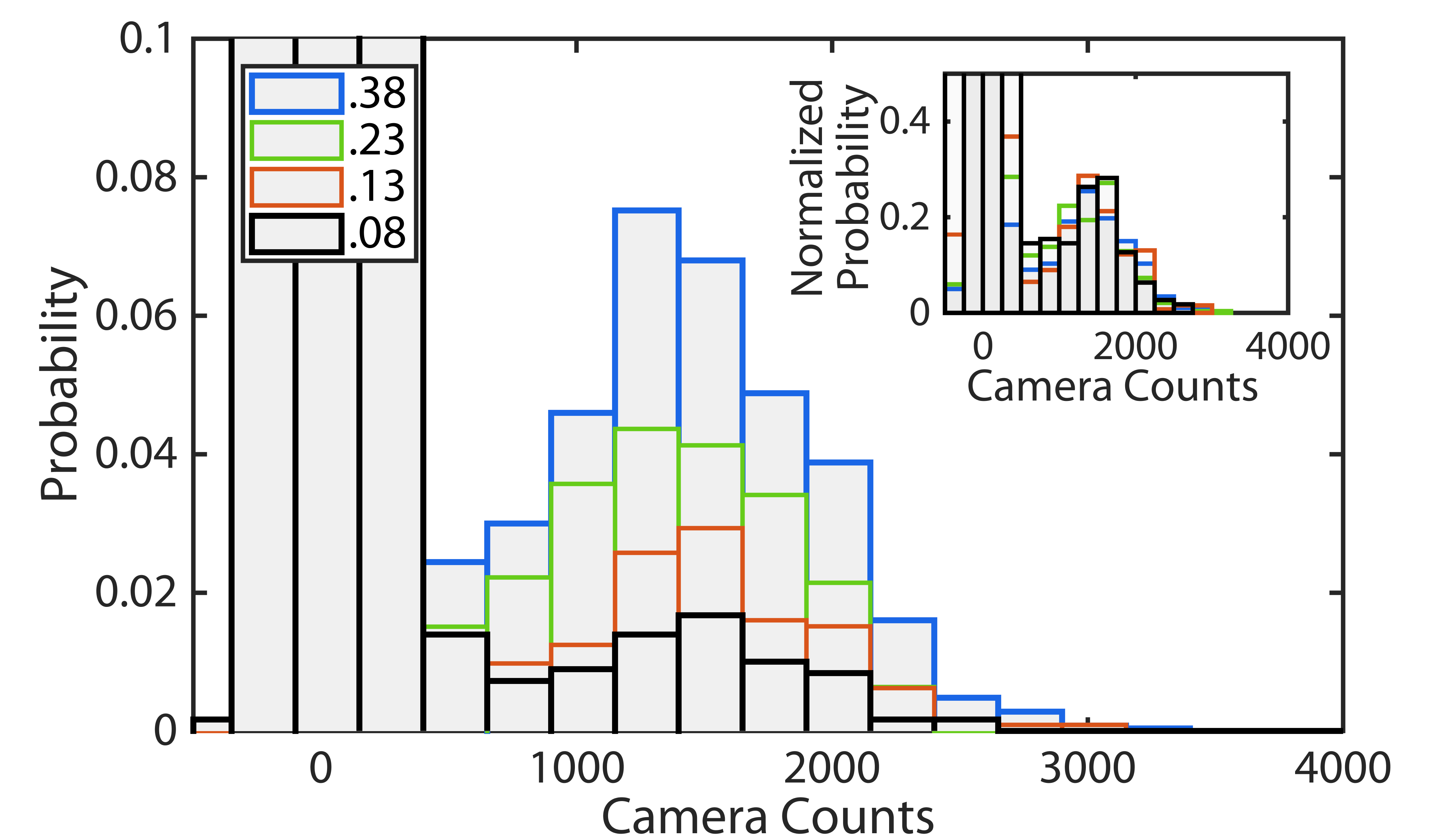}
\label{hist} 
\caption{\textbf{Histograms for Single Molecules}. Histograms with various tweezer loading fractions as indicated by the legend. Inset: Histogram normalized by camera counts under the secondary feature.}
\end{figure}

To ensure that at most one molecule is contained in each trap, we make use of light-assisted collisions in the presence of near-resonant light to ``clean out'' multiple occupancies, as is routinely done in atomic tweezer experiments~\cite{schlosser01}. Inelastic loss from light-assisted collisions leads to a collisional blockade mechanism that ensures that the occupation of each tweezer trap is either one or zero~\cite{schlosser02}. This provides a clean starting point, where uniform defect-free arrays of single molecules can be created simply by rearranging the positions of occupied traps.

In order to induce possible light-assisted collisions, we leave the $\Lambda$-cooling light on for an additional $5\,\text{ms}$ after loading. To determine whether multiple occupancies occur, we use background-subtracted single-shot images to produce histograms of photon counts in each tweezer trap. These histograms reveal a peak centered at zero counts corresponding to zero molecules, and a secondary feature with a peak centered around 1500 camera counts. To verify that this secondary feature corresponds to single molecules, we progressively reduce the loading rate into the tweezers by reducing the initial MOT number. As shown in Fig.~3, the center of the second feature remains unchanged, while its height decreases. When normalized to the area under the secondary feature, the second feature overlaps in all the histograms. This demonstrates that at most one molecule is present in each trap. If there were more than one molecule, the center of the secondary feature would move towards the zero molecule feature as the average number in the tweezer is reduced. 

\begin{figure}[t]
\centering
\includegraphics[scale=.42]{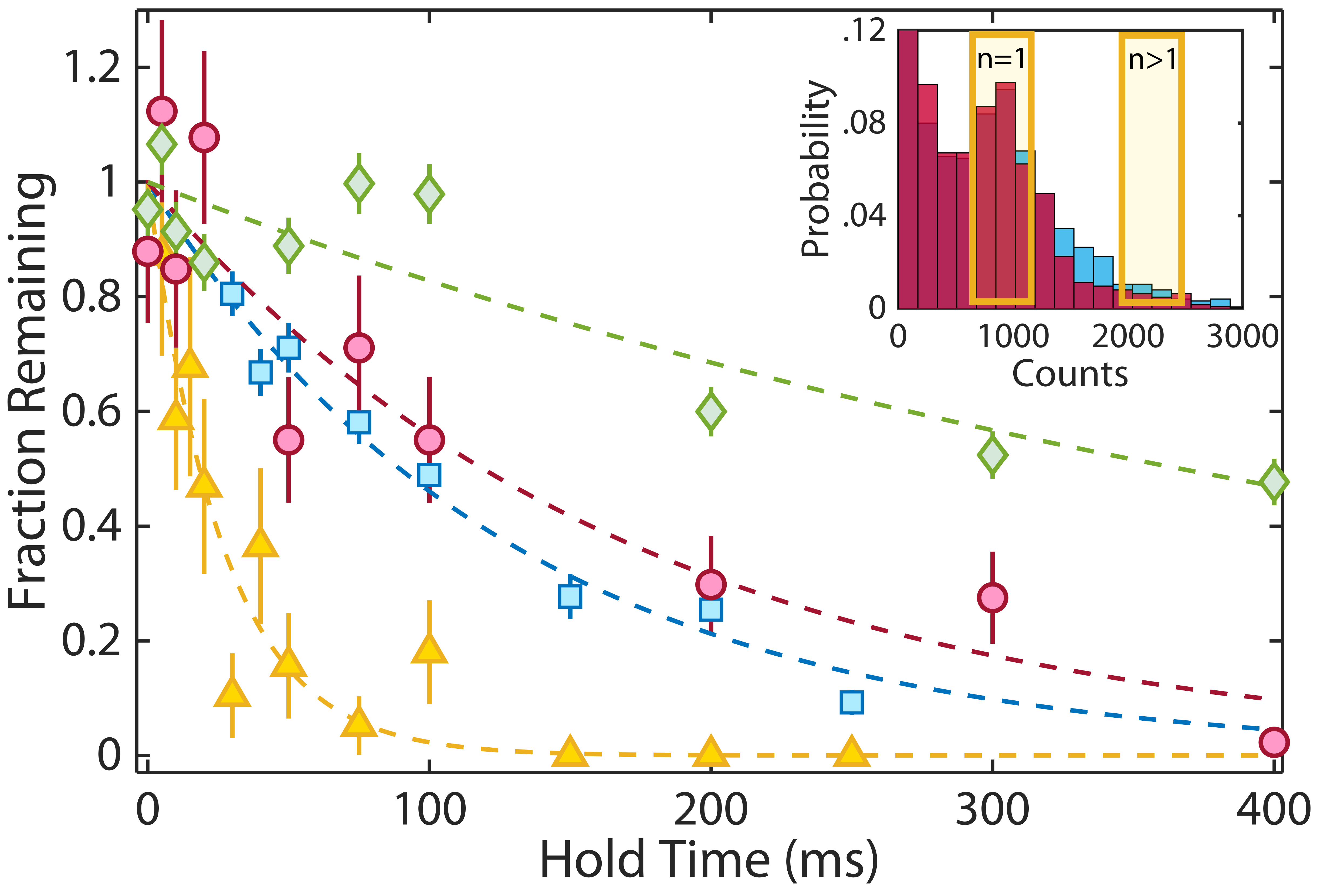}
\label{collisions} 
\caption{\textbf{Molecular Collisions in the Tweezer} Loss curves for a single molecule (green diamonds) and two molecules (red circles) in the absence of light. The single molecule lifetime of $\tau=530$\,ms and the two molecule lifetime is $\tau=180$\,ms. Loss curves for a single molecule (blue squares) and two molecules (yellow triangles) in the presence of $\Lambda$-cooling light, reflect the imaging lifetime and light induced collisional loss rate, respectively. The imaging lifetime is $\tau=130$\,ms and the light induced collision lifetime is $\tau=26$\,ms. Inset: Histograms of camera counts at short (blue) and long (red) hold times. The highlighted region $n=1$ ($n>1$) denotes the range used to compute the fraction remaining for single (double) molecules.} 
\end{figure}

These histograms also allow us to determine the detection fidelity for single molecules in a single shot. For each image, a tweezer trap is determined to be occupied if the number of photon counts exceeds a certain threshold. Due to technical noise and background light, multiple photons per molecule are needed to make a determination, and higher fidelities are obtained with higher number of collected photons. Although the number of photons emitted can be increased with longer imaging durations, durations longer than the imaging lifetime of $\sim 100$\,ms do not help since it leads to increased background light. After optimizing the imaging parameters and duration, we reach a detection fidelity of 92\% at an optimal exposure of 30\,ms exposure.

To measure the rate of light-assisted collisions directly we enlarge the size of the optical tweezer trap from $2.3\,\mu\text{m}$ to $3.6\,\mu\text{m}$ in order to lower the trapped densities and hence the collision rates. Lower collision rates provide a time window where multiple molecules can be laser cooled into the optical tweezer trap before collisional losses set in. The decay of multiple molecules is then studied by comparing histograms obtained after various $\Lambda$-cooling times. For short times, the histograms (Fig.~4 Inset) display a long tail extending beyond the single molecule peak, suggesting that multiple molecules are loaded. This long tail decays with longer $\Lambda$-cooling times, at a rate much quicker than that of the single molecule peak. To quantify the loss rate, we threshold the histogram above the single molecule feature (Fig.~4 inset), and measure the fraction above this threshold, $f_{n>1}$, which serves as a proxy for probability of loading more than one molecule. As shown in Fig.~4, the decay of $f_{n>1}$ as a function $\Lambda$-cooling time yields a $1/e$ lifetime of $26\,$ms. By comparison, the lifetime of single molecules in the presence of $\Lambda$-cooling time is $130\,$ms. This demonstrates that collisions are indeed occurring. 

To determine whether collisions are occuring in the absence of light, we perform the same measurements by holding the molecules in the dark. While the single molecule lifetime is $530$\,ms (Fig.~4), $f_{n>1}$ has a lifetime of $180$\,ms, which indicates that collisions are also present in the absence of light albeit at a much lower rate. Since the collisional loss rate scales with molecular density, with the smaller $2.3\,\mu\text{m}$ used in the array $\Lambda$-cooling of a few ms is sufficient to induce light-assisted collisions. This is consistent with the observed absence of multiple molecules in the smaller trap (Fig.~3). Assuming that the decay of $f_{n>1}$ is primarily from two molecules, we obtain a light induced collision rate of $\gamma=2(1)\times10^{-8}$cm$^{-3}$s$^{-1}$, corresponing to a cross section of $\sigma=7(3)\times10^{-10}$cm$^{2}$. This is similar to that measured for Rb atoms in optical tweezers~\cite{Fuhrmanek12}. Since light-induced collisions arise from dipolar interactions that result from an electric dipole moment induced by near-resonant light~\cite{schlosser01,schlosser02}, which have similar sizes in both atomic and molecular systems of size $\sim 1\,\text{Debye}$, similar rates to atomic systems may be expected. We note that the light-assisted collisional cross-section indicates a density limit of $\sim 10^{11}\,\text{cm}^{-3}$ for the typical ms timescales for laser cooling. In the absence of light, the fitted loss rate is $\gamma=2(1)\times10^{-9}$~cm$^{-3}$s$^{-1}$, corresponding to a cross section of $\sigma=1.0(5)\times10^{-10}$cm$^{2}$. There are multiple possible loss mechanisms, ranging from hyperfine and rotational relaxation to long-lived complex formation to simple elastic collisional loss. Detailed characterization awaits future work. Controlled merging of singly-occupied tweezer traps and internal state preparation~\cite{blackmore18} in the future would provide a clean platform for such collisional studies~\cite{Liu18}.

In conclusion, we have loaded and detected with high fidelity an array of single molecules trapped in optical tweezers. We have observed for the first time ground electronic state collisions and light-induced collisions for laser cooled molecules, which sets a density limit to $\Lambda$-cooling of CaF. By applying microwaves or DC electric fields, long-range dipolar interactions between single molecules could be engineered in future experiments. This paves the way for molecular tweezer arrays to be a quantum simulation and qubit platform with efficient state preparation and detection due to the inherent high signal from photon cycling, as in atoms. The methods we have developed in this work could also be extended to other laser coolable molecules, including polyatomic ones, opening up a variety of future applications ranging from precision measurements~\cite{Lim2018,Kozyryev17,carr09} to ultracold chemistry~\cite{krems08, carr09, kozyryev16}.

This work was supported by the NSF. We thank members of the CUA atom array experiment for useful discussions and code to run the USRP. LWC acknowledges supports from the MPHQ.

\bibliography{tweezerbib} 

\begin{thebibliography}{36}%
\makeatletter
\providecommand \@ifxundefined [1]{%
 \@ifx{#1\undefined}
}%
\providecommand \@ifnum [1]{%
 \ifnum #1\expandafter \@firstoftwo
 \else \expandafter \@secondoftwo
 \fi
}%
\providecommand \@ifx [1]{%
 \ifx #1\expandafter \@firstoftwo
 \else \expandafter \@secondoftwo
 \fi
}%
\providecommand \natexlab [1]{#1}%
\providecommand \enquote  [1]{``#1''}%
\providecommand \bibnamefont  [1]{#1}%
\providecommand \bibfnamefont [1]{#1}%
\providecommand \citenamefont [1]{#1}%
\providecommand \href@noop [0]{\@secondoftwo}%
\providecommand \href [0]{\begingroup \@sanitize@url \@href}%
\providecommand \@href[1]{\@@startlink{#1}\@@href}%
\providecommand \@@href[1]{\endgroup#1\@@endlink}%
\providecommand \@sanitize@url [0]{\catcode `\\12\catcode `\$12\catcode
  `\&12\catcode `\#12\catcode `\^12\catcode `\_12\catcode `\%12\relax}%
\providecommand \@@startlink[1]{}%
\providecommand \@@endlink[0]{}%
\providecommand \url  [0]{\begingroup\@sanitize@url \@url }%
\providecommand \@url [1]{\endgroup\@href {#1}{\urlprefix }}%
\providecommand \urlprefix  [0]{URL }%
\providecommand \Eprint [0]{\href }%
\providecommand \doibase [0]{http://dx.doi.org/}%
\providecommand \selectlanguage [0]{\@gobble}%
\providecommand \bibinfo  [0]{\@secondoftwo}%
\providecommand \bibfield  [0]{\@secondoftwo}%
\providecommand \translation [1]{[#1]}%
\providecommand \BibitemOpen [0]{}%
\providecommand \bibitemStop [0]{}%
\providecommand \bibitemNoStop [0]{.\EOS\space}%
\providecommand \EOS [0]{\spacefactor3000\relax}%
\providecommand \BibitemShut  [1]{\csname bibitem#1\endcsname}%
\let\auto@bib@innerbib\@empty
%</preamble>
\bibitem [{\citenamefont {Carr}\ \emph {et~al.}(2009)\citenamefont {Carr},
  \citenamefont {DeMille}, \citenamefont {Krems},\ and\ \citenamefont
  {Ye}}]{carr09}%
  \BibitemOpen
  \bibfield  {author} {\bibinfo {author} {\bibfnamefont {L.~D.}\ \bibnamefont
  {Carr}}, \bibinfo {author} {\bibfnamefont {D.}~\bibnamefont {DeMille}},
  \bibinfo {author} {\bibfnamefont {R.~V.}\ \bibnamefont {Krems}}, \ and\
  \bibinfo {author} {\bibfnamefont {J.}~\bibnamefont {Ye}},\ }\href {\doibase
  10.1088/1367-2630/11/5/055049} {\bibfield  {journal} {\bibinfo  {journal}
  {New Journal of Physics}\ }\textbf {\bibinfo {volume} {11}},\ \bibinfo
  {pages} {055049} (\bibinfo {year} {2009})}\BibitemShut {NoStop}%
\bibitem [{\citenamefont {DeMille}(2002)}]{demille02qi}%
  \BibitemOpen
  \bibfield  {author} {\bibinfo {author} {\bibfnamefont {D.}~\bibnamefont
  {DeMille}},\ }\href@noop {} {\bibfield  {journal} {\bibinfo  {journal}
  {Physical Review Letters}\ }\textbf {\bibinfo {volume} {88}},\ \bibinfo
  {pages} {067901} (\bibinfo {year} {2002})}\BibitemShut {NoStop}%
\bibitem [{\citenamefont {Yelin}\ \emph {et~al.}(2006)\citenamefont {Yelin},
  \citenamefont {Kirby},\ and\ \citenamefont {C{\^{o}}t{\'{e}}}}]{yelin06}%
  \BibitemOpen
  \bibfield  {author} {\bibinfo {author} {\bibfnamefont {S.~F.}\ \bibnamefont
  {Yelin}}, \bibinfo {author} {\bibfnamefont {K.}~\bibnamefont {Kirby}}, \ and\
  \bibinfo {author} {\bibfnamefont {R.}~\bibnamefont {C{\^{o}}t{\'{e}}}},\
  }\href {\doibase 10.1103/physreva.74.050301} {\bibfield  {journal} {\bibinfo
  {journal} {Physical Review A}\ }\textbf {\bibinfo {volume} {74}},\ \bibinfo
  {pages} {050301} (\bibinfo {year} {2006})}\BibitemShut {NoStop}%
\bibitem [{\citenamefont {Karra}\ \emph {et~al.}(2016)\citenamefont {Karra},
  \citenamefont {Sharma}, \citenamefont {Friedrich}, \citenamefont {Kais},\
  and\ \citenamefont {Herschbach}}]{karra16}%
  \BibitemOpen
  \bibfield  {author} {\bibinfo {author} {\bibfnamefont {M.}~\bibnamefont
  {Karra}}, \bibinfo {author} {\bibfnamefont {K.}~\bibnamefont {Sharma}},
  \bibinfo {author} {\bibfnamefont {B.}~\bibnamefont {Friedrich}}, \bibinfo
  {author} {\bibfnamefont {S.}~\bibnamefont {Kais}}, \ and\ \bibinfo {author}
  {\bibfnamefont {D.}~\bibnamefont {Herschbach}},\ }\href {\doibase
  10.1063/1.4942928} {\bibfield  {journal} {\bibinfo  {journal} {The Journal of
  Chemical Physics}\ }\textbf {\bibinfo {volume} {144}},\ \bibinfo {pages}
  {094301} (\bibinfo {year} {2016})}\BibitemShut {NoStop}%
\bibitem [{\citenamefont {Ni}\ \emph {et~al.}(2018)\citenamefont {Ni},
  \citenamefont {Rosenband},\ and\ \citenamefont {Grimes}}]{ni18}%
  \BibitemOpen
  \bibfield  {author} {\bibinfo {author} {\bibfnamefont {K.-K.}\ \bibnamefont
  {Ni}}, \bibinfo {author} {\bibfnamefont {T.}~\bibnamefont {Rosenband}}, \
  and\ \bibinfo {author} {\bibfnamefont {D.~D.}\ \bibnamefont {Grimes}},\
  }\href {\doibase 10.1039/c8sc02355g} {\bibfield  {journal} {\bibinfo
  {journal} {Chemical Science}\ }\textbf {\bibinfo {volume} {9}},\ \bibinfo
  {pages} {6830} (\bibinfo {year} {2018})}\BibitemShut {NoStop}%
\bibitem [{\citenamefont {Blackmore}\ \emph {et~al.}(2018)\citenamefont
  {Blackmore}, \citenamefont {Caldwell}, \citenamefont {Gregory}, \citenamefont
  {Bridge}, \citenamefont {Sawant}, \citenamefont {Aldegunde}, \citenamefont
  {Mur-Petit}, \citenamefont {Jaksch}, \citenamefont {Hutson}, \citenamefont
  {Sauer}, \citenamefont {Tarbutt},\ and\ \citenamefont
  {Cornish}}]{blackmore18}%
  \BibitemOpen
  \bibfield  {author} {\bibinfo {author} {\bibfnamefont {J.~A.}\ \bibnamefont
  {Blackmore}}, \bibinfo {author} {\bibfnamefont {L.}~\bibnamefont {Caldwell}},
  \bibinfo {author} {\bibfnamefont {P.~D.}\ \bibnamefont {Gregory}}, \bibinfo
  {author} {\bibfnamefont {E.~M.}\ \bibnamefont {Bridge}}, \bibinfo {author}
  {\bibfnamefont {R.}~\bibnamefont {Sawant}}, \bibinfo {author} {\bibfnamefont
  {J.}~\bibnamefont {Aldegunde}}, \bibinfo {author} {\bibfnamefont
  {J.}~\bibnamefont {Mur-Petit}}, \bibinfo {author} {\bibfnamefont
  {D.}~\bibnamefont {Jaksch}}, \bibinfo {author} {\bibfnamefont {J.~M.}\
  \bibnamefont {Hutson}}, \bibinfo {author} {\bibfnamefont {B.~E.}\
  \bibnamefont {Sauer}}, \bibinfo {author} {\bibfnamefont {M.~R.}\ \bibnamefont
  {Tarbutt}}, \ and\ \bibinfo {author} {\bibfnamefont {S.~L.}\ \bibnamefont
  {Cornish}},\ }\href@noop {} {\bibfield  {journal} {\bibinfo  {journal}
  {Quantum Science and Technology}\ }\textbf {\bibinfo {volume} {4}},\ \bibinfo
  {pages} {014010} (\bibinfo {year} {2018})}\BibitemShut {NoStop}%
\bibitem [{\citenamefont {Micheli}\ \emph {et~al.}(2006)\citenamefont
  {Micheli}, \citenamefont {Brennen},\ and\ \citenamefont {Zoller}}]{zoller06}%
  \BibitemOpen
  \bibfield  {author} {\bibinfo {author} {\bibfnamefont {A.}~\bibnamefont
  {Micheli}}, \bibinfo {author} {\bibfnamefont {G.~K.}\ \bibnamefont
  {Brennen}}, \ and\ \bibinfo {author} {\bibfnamefont {P.}~\bibnamefont
  {Zoller}},\ }\href {\doibase 10.1038/nphys287} {\bibfield  {journal}
  {\bibinfo  {journal} {Nature Physics}\ }\textbf {\bibinfo {volume} {2}},\
  \bibinfo {pages} {341} (\bibinfo {year} {2006})}\BibitemShut {NoStop}%
\bibitem [{\citenamefont {Büchler}\ \emph {et~al.}(2007)\citenamefont
  {Büchler}, \citenamefont {Micheli},\ and\ \citenamefont
  {Zoller}}]{buechler07}%
  \BibitemOpen
  \bibfield  {author} {\bibinfo {author} {\bibfnamefont {H.~P.}\ \bibnamefont
  {Büchler}}, \bibinfo {author} {\bibfnamefont {A.}~\bibnamefont {Micheli}}, \
  and\ \bibinfo {author} {\bibfnamefont {P.}~\bibnamefont {Zoller}},\ }\href
  {\doibase 10.1038/nphys678} {\bibfield  {journal} {\bibinfo  {journal}
  {Nature Physics}\ }\textbf {\bibinfo {volume} {3}},\ \bibinfo {pages} {726}
  (\bibinfo {year} {2007})}\BibitemShut {NoStop}%
\bibitem [{\citenamefont {Pupillo}\ \emph {et~al.}(2008)\citenamefont
  {Pupillo}, \citenamefont {Griessner}, \citenamefont {Micheli}, \citenamefont
  {Ortner}, \citenamefont {Wang},\ and\ \citenamefont {Zoller}}]{pupillo08}%
  \BibitemOpen
  \bibfield  {author} {\bibinfo {author} {\bibfnamefont {G.}~\bibnamefont
  {Pupillo}}, \bibinfo {author} {\bibfnamefont {A.}~\bibnamefont {Griessner}},
  \bibinfo {author} {\bibfnamefont {A.}~\bibnamefont {Micheli}}, \bibinfo
  {author} {\bibfnamefont {M.}~\bibnamefont {Ortner}}, \bibinfo {author}
  {\bibfnamefont {D.-W.}\ \bibnamefont {Wang}}, \ and\ \bibinfo {author}
  {\bibfnamefont {P.}~\bibnamefont {Zoller}},\ }\href@noop {} {\bibfield
  {journal} {\bibinfo  {journal} {Physical Review Letters}\ }\textbf {\bibinfo
  {volume} {100}},\ \bibinfo {pages} {050402} (\bibinfo {year}
  {2008})}\BibitemShut {NoStop}%
\bibitem [{\citenamefont {Krems}(2008)}]{krems08}%
  \BibitemOpen
  \bibfield  {author} {\bibinfo {author} {\bibfnamefont {R.~V.}\ \bibnamefont
  {Krems}},\ }\href {\doibase 10.1039/b802322k} {\bibfield  {journal} {\bibinfo
   {journal} {Physical Chemistry Chemical Physics}\ }\textbf {\bibinfo {volume}
  {10}},\ \bibinfo {pages} {4079} (\bibinfo {year} {2008})}\BibitemShut
  {NoStop}%
\bibitem [{\citenamefont {Balakrishnan}(2016)}]{Bala16}%
  \BibitemOpen
  \bibfield  {author} {\bibinfo {author} {\bibfnamefont {N.}~\bibnamefont
  {Balakrishnan}},\ }\href {\doibase 10.1063/1.4964096} {\bibfield  {journal}
  {\bibinfo  {journal} {The Journal of Chemical Physics}\ }\textbf {\bibinfo
  {volume} {145}},\ \bibinfo {pages} {150901} (\bibinfo {year}
  {2016})}\BibitemShut {NoStop}%
\bibitem [{\citenamefont {Ye}\ \emph {et~al.}(2006)\citenamefont {Ye},
  \citenamefont {Blatt}, \citenamefont {Boyd}, \citenamefont {Foreman},
  \citenamefont {Hudson}, \citenamefont {Ido}, \citenamefont {Lev},
  \citenamefont {Ludlow}, \citenamefont {Sawyer}, \citenamefont {Stuhl},\ and\
  \citenamefont {Zelevinsky}}]{Ye06}%
  \BibitemOpen
  \bibfield  {author} {\bibinfo {author} {\bibfnamefont {J.}~\bibnamefont
  {Ye}}, \bibinfo {author} {\bibfnamefont {S.}~\bibnamefont {Blatt}}, \bibinfo
  {author} {\bibfnamefont {M.~M.}\ \bibnamefont {Boyd}}, \bibinfo {author}
  {\bibfnamefont {S.~M.}\ \bibnamefont {Foreman}}, \bibinfo {author}
  {\bibfnamefont {E.~R.}\ \bibnamefont {Hudson}}, \bibinfo {author}
  {\bibfnamefont {T.}~\bibnamefont {Ido}}, \bibinfo {author} {\bibfnamefont
  {B.}~\bibnamefont {Lev}}, \bibinfo {author} {\bibfnamefont {A.~D.}\
  \bibnamefont {Ludlow}}, \bibinfo {author} {\bibfnamefont {B.~C.}\
  \bibnamefont {Sawyer}}, \bibinfo {author} {\bibfnamefont {B.}~\bibnamefont
  {Stuhl}}, \ and\ \bibinfo {author} {\bibfnamefont {T.}~\bibnamefont
  {Zelevinsky}},\ }in\ \href {\doibase 10.1063/1.2400637} {\emph {\bibinfo
  {booktitle} {{AIP} Conference Proceedings}}}\ (\bibinfo  {publisher}
  {{AIP}},\ \bibinfo {year} {2006})\BibitemShut {NoStop}%
\bibitem [{\citenamefont {Barry}\ \emph {et~al.}(2014)\citenamefont {Barry},
  \citenamefont {McCarron}, \citenamefont {Norrgard}, \citenamefont
  {Steinecker},\ and\ \citenamefont {DeMille}}]{barry14}%
  \BibitemOpen
  \bibfield  {author} {\bibinfo {author} {\bibfnamefont {J.~F.}\ \bibnamefont
  {Barry}}, \bibinfo {author} {\bibfnamefont {D.~J.}\ \bibnamefont {McCarron}},
  \bibinfo {author} {\bibfnamefont {E.~B.}\ \bibnamefont {Norrgard}}, \bibinfo
  {author} {\bibfnamefont {M.~H.}\ \bibnamefont {Steinecker}}, \ and\ \bibinfo
  {author} {\bibfnamefont {D.}~\bibnamefont {DeMille}},\ }\href {\doibase
  10.1038/nature13634} {\bibfield  {journal} {\bibinfo  {journal} {Nature}\
  }\textbf {\bibinfo {volume} {512}},\ \bibinfo {pages} {286} (\bibinfo {year}
  {2014})}\BibitemShut {NoStop}%
\bibitem [{\citenamefont {Norrgard}\ \emph {et~al.}(2016)\citenamefont
  {Norrgard}, \citenamefont {McCarron}, \citenamefont {Steinecker},
  \citenamefont {Tarbutt},\ and\ \citenamefont {DeMille}}]{norrgard16RF}%
  \BibitemOpen
  \bibfield  {author} {\bibinfo {author} {\bibfnamefont {E.}~\bibnamefont
  {Norrgard}}, \bibinfo {author} {\bibfnamefont {D.}~\bibnamefont {McCarron}},
  \bibinfo {author} {\bibfnamefont {M.}~\bibnamefont {Steinecker}}, \bibinfo
  {author} {\bibfnamefont {M.}~\bibnamefont {Tarbutt}}, \ and\ \bibinfo
  {author} {\bibfnamefont {D.}~\bibnamefont {DeMille}},\ }\href@noop {}
  {\bibfield  {journal} {\bibinfo  {journal} {Phys. Rev. Lett.}\ }\textbf
  {\bibinfo {volume} {116}},\ \bibinfo {pages} {063004} (\bibinfo {year}
  {2016})}\BibitemShut {NoStop}%
\bibitem [{\citenamefont {Truppe}\ \emph {et~al.}(2017)\citenamefont {Truppe},
  \citenamefont {Williams}, \citenamefont {Hambach}, \citenamefont {Caldwell},
  \citenamefont {Fitch}, \citenamefont {Hinds}, \citenamefont {Sauer},\ and\
  \citenamefont {Tarbutt}}]{truppe17}%
  \BibitemOpen
  \bibfield  {author} {\bibinfo {author} {\bibfnamefont {S.}~\bibnamefont
  {Truppe}}, \bibinfo {author} {\bibfnamefont {H.~J.}\ \bibnamefont
  {Williams}}, \bibinfo {author} {\bibfnamefont {M.}~\bibnamefont {Hambach}},
  \bibinfo {author} {\bibfnamefont {L.}~\bibnamefont {Caldwell}}, \bibinfo
  {author} {\bibfnamefont {N.~J.}\ \bibnamefont {Fitch}}, \bibinfo {author}
  {\bibfnamefont {E.~A.}\ \bibnamefont {Hinds}}, \bibinfo {author}
  {\bibfnamefont {B.~E.}\ \bibnamefont {Sauer}}, \ and\ \bibinfo {author}
  {\bibfnamefont {M.~R.}\ \bibnamefont {Tarbutt}},\ }\href {\doibase
  10.1038/nphys4241} {\bibfield  {journal} {\bibinfo  {journal} {Nature
  Physics}\ }\textbf {\bibinfo {volume} {13}},\ \bibinfo {pages} {1173}
  (\bibinfo {year} {2017})}\BibitemShut {NoStop}%
\bibitem [{\citenamefont {Anderegg}\ \emph {et~al.}(2017)\citenamefont
  {Anderegg}, \citenamefont {Augenbraun}, \citenamefont {Chae}, \citenamefont
  {Hemmerling}, \citenamefont {Hutzler}, \citenamefont {Ravi}, \citenamefont
  {Collopy}, \citenamefont {Ye}, \citenamefont {Ketterle},\ and\ \citenamefont
  {Doyle}}]{anderegg17}%
  \BibitemOpen
  \bibfield  {author} {\bibinfo {author} {\bibfnamefont {L.}~\bibnamefont
  {Anderegg}}, \bibinfo {author} {\bibfnamefont {B.~L.}\ \bibnamefont
  {Augenbraun}}, \bibinfo {author} {\bibfnamefont {E.}~\bibnamefont {Chae}},
  \bibinfo {author} {\bibfnamefont {B.}~\bibnamefont {Hemmerling}}, \bibinfo
  {author} {\bibfnamefont {N.~R.}\ \bibnamefont {Hutzler}}, \bibinfo {author}
  {\bibfnamefont {A.}~\bibnamefont {Ravi}}, \bibinfo {author} {\bibfnamefont
  {A.}~\bibnamefont {Collopy}}, \bibinfo {author} {\bibfnamefont
  {J.}~\bibnamefont {Ye}}, \bibinfo {author} {\bibfnamefont {W.}~\bibnamefont
  {Ketterle}}, \ and\ \bibinfo {author} {\bibfnamefont {J.~M.}\ \bibnamefont
  {Doyle}},\ }\href@noop {} {\bibfield  {journal} {\bibinfo  {journal}
  {Physical Review Letters}\ }\textbf {\bibinfo {volume} {119}} (\bibinfo
  {year} {2017})}\BibitemShut {NoStop}%
\bibitem [{\citenamefont {Collopy}\ \emph {et~al.}(2018)\citenamefont
  {Collopy}, \citenamefont {Ding}, \citenamefont {Wu}, \citenamefont
  {Finneran}, \citenamefont {Anderegg}, \citenamefont {Augenbraun},
  \citenamefont {Doyle},\ and\ \citenamefont {Ye}}]{collopy18}%
  \BibitemOpen
  \bibfield  {author} {\bibinfo {author} {\bibfnamefont {A.~L.}\ \bibnamefont
  {Collopy}}, \bibinfo {author} {\bibfnamefont {S.}~\bibnamefont {Ding}},
  \bibinfo {author} {\bibfnamefont {Y.}~\bibnamefont {Wu}}, \bibinfo {author}
  {\bibfnamefont {I.~A.}\ \bibnamefont {Finneran}}, \bibinfo {author}
  {\bibfnamefont {L.}~\bibnamefont {Anderegg}}, \bibinfo {author}
  {\bibfnamefont {B.~L.}\ \bibnamefont {Augenbraun}}, \bibinfo {author}
  {\bibfnamefont {J.~M.}\ \bibnamefont {Doyle}}, \ and\ \bibinfo {author}
  {\bibfnamefont {J.}~\bibnamefont {Ye}},\ }\href@noop {} {\bibfield  {journal}
  {\bibinfo  {journal} {Physical Review Letters}\ }\textbf {\bibinfo {volume}
  {121}} (\bibinfo {year} {2018})}\BibitemShut {NoStop}%
\bibitem [{\citenamefont {Williams}\ \emph {et~al.}(2018)\citenamefont
  {Williams}, \citenamefont {Caldwell}, \citenamefont {Fitch}, \citenamefont
  {Truppe}, \citenamefont {Rodewald}, \citenamefont {Hinds}, \citenamefont
  {Sauer},\ and\ \citenamefont {Tarbutt}}]{williams18}%
  \BibitemOpen
  \bibfield  {author} {\bibinfo {author} {\bibfnamefont {H.}~\bibnamefont
  {Williams}}, \bibinfo {author} {\bibfnamefont {L.}~\bibnamefont {Caldwell}},
  \bibinfo {author} {\bibfnamefont {N.}~\bibnamefont {Fitch}}, \bibinfo
  {author} {\bibfnamefont {S.}~\bibnamefont {Truppe}}, \bibinfo {author}
  {\bibfnamefont {J.}~\bibnamefont {Rodewald}}, \bibinfo {author}
  {\bibfnamefont {E.}~\bibnamefont {Hinds}}, \bibinfo {author} {\bibfnamefont
  {B.}~\bibnamefont {Sauer}}, \ and\ \bibinfo {author} {\bibfnamefont
  {M.}~\bibnamefont {Tarbutt}},\ }\href@noop {} {\bibfield  {journal} {\bibinfo
   {journal} {Physical Review Letters}\ }\textbf {\bibinfo {volume} {120}}
  (\bibinfo {year} {2018})}\BibitemShut {NoStop}%
\bibitem [{\citenamefont {McCarron}\ \emph {et~al.}(2018)\citenamefont
  {McCarron}, \citenamefont {Steinecker}, \citenamefont {Zhu},\ and\
  \citenamefont {DeMille}}]{McCarron18}%
  \BibitemOpen
  \bibfield  {author} {\bibinfo {author} {\bibfnamefont {D.}~\bibnamefont
  {McCarron}}, \bibinfo {author} {\bibfnamefont {M.}~\bibnamefont
  {Steinecker}}, \bibinfo {author} {\bibfnamefont {Y.}~\bibnamefont {Zhu}}, \
  and\ \bibinfo {author} {\bibfnamefont {D.}~\bibnamefont {DeMille}},\
  }\href@noop {} {\bibfield  {journal} {\bibinfo  {journal} {Physical Review
  Letters}\ }\textbf {\bibinfo {volume} {121}} (\bibinfo {year}
  {2018})}\BibitemShut {NoStop}%
\bibitem [{\citenamefont {Anderegg}\ \emph {et~al.}(2018)\citenamefont
  {Anderegg}, \citenamefont {Augenbraun}, \citenamefont {Bao}, \citenamefont
  {Burchesky}, \citenamefont {Cheuk}, \citenamefont {Ketterle},\ and\
  \citenamefont {Doyle}}]{anderegg18}%
  \BibitemOpen
  \bibfield  {author} {\bibinfo {author} {\bibfnamefont {L.}~\bibnamefont
  {Anderegg}}, \bibinfo {author} {\bibfnamefont {B.~L.}\ \bibnamefont
  {Augenbraun}}, \bibinfo {author} {\bibfnamefont {Y.}~\bibnamefont {Bao}},
  \bibinfo {author} {\bibfnamefont {S.}~\bibnamefont {Burchesky}}, \bibinfo
  {author} {\bibfnamefont {L.~W.}\ \bibnamefont {Cheuk}}, \bibinfo {author}
  {\bibfnamefont {W.}~\bibnamefont {Ketterle}}, \ and\ \bibinfo {author}
  {\bibfnamefont {J.~M.}\ \bibnamefont {Doyle}},\ }\href {\doibase
  10.1038/s41567-018-0191-z} {\bibfield  {journal} {\bibinfo  {journal} {Nature
  Physics}\ }\textbf {\bibinfo {volume} {14}},\ \bibinfo {pages} {890}
  (\bibinfo {year} {2018})}\BibitemShut {NoStop}%
\bibitem [{\citenamefont {Cheuk}\ \emph {et~al.}(2018)\citenamefont {Cheuk},
  \citenamefont {Anderegg}, \citenamefont {Augenbraun}, \citenamefont {Bao},
  \citenamefont {Burchesky}, \citenamefont {Ketterle},\ and\ \citenamefont
  {Doyle}}]{cheuk18}%
  \BibitemOpen
  \bibfield  {author} {\bibinfo {author} {\bibfnamefont {L.~W.}\ \bibnamefont
  {Cheuk}}, \bibinfo {author} {\bibfnamefont {L.}~\bibnamefont {Anderegg}},
  \bibinfo {author} {\bibfnamefont {B.~L.}\ \bibnamefont {Augenbraun}},
  \bibinfo {author} {\bibfnamefont {Y.}~\bibnamefont {Bao}}, \bibinfo {author}
  {\bibfnamefont {S.}~\bibnamefont {Burchesky}}, \bibinfo {author}
  {\bibfnamefont {W.}~\bibnamefont {Ketterle}}, \ and\ \bibinfo {author}
  {\bibfnamefont {J.~M.}\ \bibnamefont {Doyle}},\ }\href
  {http://adsabs.harvard.edu/abs/2018PhRvL.121h3201C} {\bibfield  {journal}
  {\bibinfo  {journal} {Physical Review Letters}\ }\textbf {\bibinfo {volume}
  {121}},\ \bibinfo {pages} {083201} (\bibinfo {year} {2018})}\BibitemShut
  {NoStop}%
\bibitem [{\citenamefont {Marco}\ \emph {et~al.}()\citenamefont {Marco},
  \citenamefont {Valtolina}, \citenamefont {Matsuda}, \citenamefont {Tobias},
  \citenamefont {Covey},\ and\ \citenamefont {Ye}}]{Marco18}%
  \BibitemOpen
  \bibfield  {author} {\bibinfo {author} {\bibfnamefont {L.~D.}\ \bibnamefont
  {Marco}}, \bibinfo {author} {\bibfnamefont {G.}~\bibnamefont {Valtolina}},
  \bibinfo {author} {\bibfnamefont {K.}~\bibnamefont {Matsuda}}, \bibinfo
  {author} {\bibfnamefont {W.~G.}\ \bibnamefont {Tobias}}, \bibinfo {author}
  {\bibfnamefont {J.~P.}\ \bibnamefont {Covey}}, \ and\ \bibinfo {author}
  {\bibfnamefont {J.}~\bibnamefont {Ye}},\ }\href@noop {} {\ }\Eprint
  {http://arxiv.org/abs/http://arxiv.org/abs/1808.00028v1}
  {http://arxiv.org/abs/1808.00028v1} \BibitemShut {NoStop}%
\bibitem [{\citenamefont {Schlosser}\ \emph {et~al.}(2001)\citenamefont
  {Schlosser}, \citenamefont {Reymond}, \citenamefont {Protsenko},\ and\
  \citenamefont {Grangier}}]{schlosser01}%
  \BibitemOpen
  \bibfield  {author} {\bibinfo {author} {\bibfnamefont {N.}~\bibnamefont
  {Schlosser}}, \bibinfo {author} {\bibfnamefont {G.}~\bibnamefont {Reymond}},
  \bibinfo {author} {\bibfnamefont {I.}~\bibnamefont {Protsenko}}, \ and\
  \bibinfo {author} {\bibfnamefont {P.}~\bibnamefont {Grangier}},\ }\href
  {\doibase 10.1038/35082512} {\bibfield  {journal} {\bibinfo  {journal}
  {Nature}\ }\textbf {\bibinfo {volume} {411}},\ \bibinfo {pages} {1024}
  (\bibinfo {year} {2001})}\BibitemShut {NoStop}%
\bibitem [{\citenamefont {Yavuz}\ \emph {et~al.}(2006)\citenamefont {Yavuz},
  \citenamefont {Kulatunga}, \citenamefont {Urban}, \citenamefont {Johnson},
  \citenamefont {Proite}, \citenamefont {Henage}, \citenamefont {Walker},\ and\
  \citenamefont {Saffman}}]{Saffman06}%
  \BibitemOpen
  \bibfield  {author} {\bibinfo {author} {\bibfnamefont {D.~D.}\ \bibnamefont
  {Yavuz}}, \bibinfo {author} {\bibfnamefont {P.~B.}\ \bibnamefont
  {Kulatunga}}, \bibinfo {author} {\bibfnamefont {E.}~\bibnamefont {Urban}},
  \bibinfo {author} {\bibfnamefont {T.~A.}\ \bibnamefont {Johnson}}, \bibinfo
  {author} {\bibfnamefont {N.}~\bibnamefont {Proite}}, \bibinfo {author}
  {\bibfnamefont {T.}~\bibnamefont {Henage}}, \bibinfo {author} {\bibfnamefont
  {T.~G.}\ \bibnamefont {Walker}}, \ and\ \bibinfo {author} {\bibfnamefont
  {M.}~\bibnamefont {Saffman}},\ }\href@noop {} {\bibfield  {journal} {\bibinfo
   {journal} {Physical Review Letters}\ }\textbf {\bibinfo {volume} {96}}
  (\bibinfo {year} {2006})}\BibitemShut {NoStop}%
\bibitem [{\citenamefont {Kaufman}\ \emph {et~al.}(2012)\citenamefont
  {Kaufman}, \citenamefont {Lester},\ and\ \citenamefont {Regal}}]{Kaufman12}%
  \BibitemOpen
  \bibfield  {author} {\bibinfo {author} {\bibfnamefont {A.~M.}\ \bibnamefont
  {Kaufman}}, \bibinfo {author} {\bibfnamefont {B.~J.}\ \bibnamefont {Lester}},
  \ and\ \bibinfo {author} {\bibfnamefont {C.~A.}\ \bibnamefont {Regal}},\
  }\href@noop {} {\bibfield  {journal} {\bibinfo  {journal} {Physical Review
  X}\ }\textbf {\bibinfo {volume} {2}} (\bibinfo {year} {2012})}\BibitemShut
  {NoStop}%
\bibitem [{\citenamefont {Norcia}\ \emph {et~al.}(2018)\citenamefont {Norcia},
  \citenamefont {Young},\ and\ \citenamefont {Kaufman}}]{Kaufman18}%
  \BibitemOpen
  \bibfield  {author} {\bibinfo {author} {\bibfnamefont {M.}~\bibnamefont
  {Norcia}}, \bibinfo {author} {\bibfnamefont {A.}~\bibnamefont {Young}}, \
  and\ \bibinfo {author} {\bibfnamefont {A.}~\bibnamefont {Kaufman}},\
  }\href@noop {} {\bibfield  {journal} {\bibinfo  {journal} {Physical Review
  X}\ }\textbf {\bibinfo {volume} {8}} (\bibinfo {year} {2018})}\BibitemShut
  {NoStop}%
\bibitem [{\citenamefont {Cooper}\ \emph {et~al.}(2018)\citenamefont {Cooper},
  \citenamefont {Covey}, \citenamefont {Madjarov}, \citenamefont {Porsev},
  \citenamefont {Safronova},\ and\ \citenamefont {Endres}}]{cooper18}%
  \BibitemOpen
  \bibfield  {author} {\bibinfo {author} {\bibfnamefont {A.}~\bibnamefont
  {Cooper}}, \bibinfo {author} {\bibfnamefont {J.~P.}\ \bibnamefont {Covey}},
  \bibinfo {author} {\bibfnamefont {I.~S.}\ \bibnamefont {Madjarov}}, \bibinfo
  {author} {\bibfnamefont {S.~G.}\ \bibnamefont {Porsev}}, \bibinfo {author}
  {\bibfnamefont {M.~S.}\ \bibnamefont {Safronova}}, \ and\ \bibinfo {author}
  {\bibfnamefont {M.}~\bibnamefont {Endres}},\ }\href@noop {} {\bibfield
  {journal} {\bibinfo  {journal} {Physical Review X}\ }\textbf {\bibinfo
  {volume} {8}} (\bibinfo {year} {2018})}\BibitemShut {NoStop}%
\bibitem [{\citenamefont {Saskin}\ \emph {et~al.}()\citenamefont {Saskin},
  \citenamefont {Wilson}, \citenamefont {Grinkemeyer},\ and\ \citenamefont
  {Thompson}}]{Saskin18}%
  \BibitemOpen
  \bibfield  {author} {\bibinfo {author} {\bibfnamefont {S.}~\bibnamefont
  {Saskin}}, \bibinfo {author} {\bibfnamefont {J.}~\bibnamefont {Wilson}},
  \bibinfo {author} {\bibfnamefont {B.}~\bibnamefont {Grinkemeyer}}, \ and\
  \bibinfo {author} {\bibfnamefont {J.}~\bibnamefont {Thompson}},\ }\href@noop
  {} {\ }\Eprint {http://arxiv.org/abs/http://arxiv.org/abs/1810.10517v1}
  {http://arxiv.org/abs/1810.10517v1} \BibitemShut {NoStop}%
\bibitem [{\citenamefont {Barredo}\ \emph {et~al.}(2016)\citenamefont
  {Barredo}, \citenamefont {de~L{\'{e}}s{\'{e}}leuc}, \citenamefont {Lienhard},
  \citenamefont {Lahaye},\ and\ \citenamefont {Browaeys}}]{barredo16array}%
  \BibitemOpen
  \bibfield  {author} {\bibinfo {author} {\bibfnamefont {D.}~\bibnamefont
  {Barredo}}, \bibinfo {author} {\bibfnamefont {S.}~\bibnamefont
  {de~L{\'{e}}s{\'{e}}leuc}}, \bibinfo {author} {\bibfnamefont
  {V.}~\bibnamefont {Lienhard}}, \bibinfo {author} {\bibfnamefont
  {T.}~\bibnamefont {Lahaye}}, \ and\ \bibinfo {author} {\bibfnamefont
  {A.}~\bibnamefont {Browaeys}},\ }\href {\doibase 10.1126/science.aah3778}
  {\bibfield  {journal} {\bibinfo  {journal} {Science}\ }\textbf {\bibinfo
  {volume} {354}},\ \bibinfo {pages} {1021} (\bibinfo {year}
  {2016})}\BibitemShut {NoStop}%
\bibitem [{\citenamefont {Endres}\ \emph {et~al.}(2016)\citenamefont {Endres},
  \citenamefont {Bernien}, \citenamefont {Keesling}, \citenamefont {Levine},
  \citenamefont {Anschuetz}, \citenamefont {Krajenbrink}, \citenamefont
  {Senko}, \citenamefont {Vuletic}, \citenamefont {Greiner},\ and\
  \citenamefont {Lukin}}]{lukin16array}%
  \BibitemOpen
  \bibfield  {author} {\bibinfo {author} {\bibfnamefont {M.}~\bibnamefont
  {Endres}}, \bibinfo {author} {\bibfnamefont {H.}~\bibnamefont {Bernien}},
  \bibinfo {author} {\bibfnamefont {A.}~\bibnamefont {Keesling}}, \bibinfo
  {author} {\bibfnamefont {H.}~\bibnamefont {Levine}}, \bibinfo {author}
  {\bibfnamefont {E.~R.}\ \bibnamefont {Anschuetz}}, \bibinfo {author}
  {\bibfnamefont {A.}~\bibnamefont {Krajenbrink}}, \bibinfo {author}
  {\bibfnamefont {C.}~\bibnamefont {Senko}}, \bibinfo {author} {\bibfnamefont
  {V.}~\bibnamefont {Vuletic}}, \bibinfo {author} {\bibfnamefont
  {M.}~\bibnamefont {Greiner}}, \ and\ \bibinfo {author} {\bibfnamefont
  {M.~D.}\ \bibnamefont {Lukin}},\ }\href {\doibase 10.1126/science.aah3752}
  {\bibfield  {journal} {\bibinfo  {journal} {Science}\ }\textbf {\bibinfo
  {volume} {354}},\ \bibinfo {pages} {1024} (\bibinfo {year}
  {2016})}\BibitemShut {NoStop}%
\bibitem [{\citenamefont {Liu}\ \emph {et~al.}(2018)\citenamefont {Liu},
  \citenamefont {Hood}, \citenamefont {Yu}, \citenamefont {Zhang},
  \citenamefont {Hutzler}, \citenamefont {Rosenband},\ and\ \citenamefont
  {Ni}}]{Liu18}%
  \BibitemOpen
  \bibfield  {author} {\bibinfo {author} {\bibfnamefont {L.~R.}\ \bibnamefont
  {Liu}}, \bibinfo {author} {\bibfnamefont {J.~D.}\ \bibnamefont {Hood}},
  \bibinfo {author} {\bibfnamefont {Y.}~\bibnamefont {Yu}}, \bibinfo {author}
  {\bibfnamefont {J.~T.}\ \bibnamefont {Zhang}}, \bibinfo {author}
  {\bibfnamefont {N.~R.}\ \bibnamefont {Hutzler}}, \bibinfo {author}
  {\bibfnamefont {T.}~\bibnamefont {Rosenband}}, \ and\ \bibinfo {author}
  {\bibfnamefont {K.-K.}\ \bibnamefont {Ni}},\ }\href {\doibase
  10.1126/science.aar7797} {\bibfield  {journal} {\bibinfo  {journal}
  {Science}\ }\textbf {\bibinfo {volume} {360}},\ \bibinfo {pages} {900}
  (\bibinfo {year} {2018})}\BibitemShut {NoStop}%
\bibitem [{\citenamefont {Schlosser}\ \emph {et~al.}(2002)\citenamefont
  {Schlosser}, \citenamefont {Reymond},\ and\ \citenamefont
  {Grangier}}]{schlosser02}%
  \BibitemOpen
  \bibfield  {author} {\bibinfo {author} {\bibfnamefont {N.}~\bibnamefont
  {Schlosser}}, \bibinfo {author} {\bibfnamefont {G.}~\bibnamefont {Reymond}},
  \ and\ \bibinfo {author} {\bibfnamefont {P.}~\bibnamefont {Grangier}},\
  }\href@noop {} {\bibfield  {journal} {\bibinfo  {journal} {Physical Review
  Letters}\ }\textbf {\bibinfo {volume} {89}} (\bibinfo {year}
  {2002})}\BibitemShut {NoStop}%
\bibitem [{\citenamefont {Fuhrmanek}\ \emph {et~al.}(2012)\citenamefont
  {Fuhrmanek}, \citenamefont {Bourgain}, \citenamefont {Sortais},\ and\
  \citenamefont {Browaeys}}]{Fuhrmanek12}%
  \BibitemOpen
  \bibfield  {author} {\bibinfo {author} {\bibfnamefont {A.}~\bibnamefont
  {Fuhrmanek}}, \bibinfo {author} {\bibfnamefont {R.}~\bibnamefont {Bourgain}},
  \bibinfo {author} {\bibfnamefont {Y.~R.~P.}\ \bibnamefont {Sortais}}, \ and\
  \bibinfo {author} {\bibfnamefont {A.}~\bibnamefont {Browaeys}},\ }\href@noop
  {} {\bibfield  {journal} {\bibinfo  {journal} {Physical Review A}\ }\textbf
  {\bibinfo {volume} {85}} (\bibinfo {year} {2012})}\BibitemShut {NoStop}%
\bibitem [{\citenamefont {Lim}\ \emph {et~al.}(2018)\citenamefont {Lim},
  \citenamefont {Almond}, \citenamefont {Trigatzis}, \citenamefont {Devlin},
  \citenamefont {Fitch}, \citenamefont {Sauer}, \citenamefont {Tarbutt},\ and\
  \citenamefont {Hinds}}]{Lim2018}%
  \BibitemOpen
  \bibfield  {author} {\bibinfo {author} {\bibfnamefont {J.}~\bibnamefont
  {Lim}}, \bibinfo {author} {\bibfnamefont {J.}~\bibnamefont {Almond}},
  \bibinfo {author} {\bibfnamefont {M.}~\bibnamefont {Trigatzis}}, \bibinfo
  {author} {\bibfnamefont {J.}~\bibnamefont {Devlin}}, \bibinfo {author}
  {\bibfnamefont {N.}~\bibnamefont {Fitch}}, \bibinfo {author} {\bibfnamefont
  {B.}~\bibnamefont {Sauer}}, \bibinfo {author} {\bibfnamefont
  {M.}~\bibnamefont {Tarbutt}}, \ and\ \bibinfo {author} {\bibfnamefont
  {E.}~\bibnamefont {Hinds}},\ }\href@noop {} {\bibfield  {journal} {\bibinfo
  {journal} {Physical Review Letters}\ }\textbf {\bibinfo {volume} {120}}
  (\bibinfo {year} {2018})}\BibitemShut {NoStop}%
\bibitem [{\citenamefont {Kozyryev}\ and\ \citenamefont
  {Hutzler}(2017)}]{Kozyryev17}%
  \BibitemOpen
  \bibfield  {author} {\bibinfo {author} {\bibfnamefont {I.}~\bibnamefont
  {Kozyryev}}\ and\ \bibinfo {author} {\bibfnamefont {N.~R.}\ \bibnamefont
  {Hutzler}},\ }\href@noop {} {\bibfield  {journal} {\bibinfo  {journal}
  {Physical Review Letters}\ }\textbf {\bibinfo {volume} {119}} (\bibinfo
  {year} {2017})}\BibitemShut {NoStop}%
\bibitem [{\citenamefont {Kozyryev}\ \emph {et~al.}(2016)\citenamefont
  {Kozyryev}, \citenamefont {Baum}, \citenamefont {Matsuda},\ and\
  \citenamefont {Doyle}}]{kozyryev16}%
  \BibitemOpen
  \bibfield  {author} {\bibinfo {author} {\bibfnamefont {I.}~\bibnamefont
  {Kozyryev}}, \bibinfo {author} {\bibfnamefont {L.}~\bibnamefont {Baum}},
  \bibinfo {author} {\bibfnamefont {K.}~\bibnamefont {Matsuda}}, \ and\
  \bibinfo {author} {\bibfnamefont {J.~M.}\ \bibnamefont {Doyle}},\ }\href
  {\doibase 10.1002/cphc.201601051} {\bibfield  {journal} {\bibinfo  {journal}
  {{ChemPhysChem}}\ }\textbf {\bibinfo {volume} {17}},\ \bibinfo {pages} {3641}
  (\bibinfo {year} {2016})}\BibitemShut {NoStop}%
\end{thebibliography}%
\end{document}